\documentclass[referee,a4paper,12pt,traditabstract]{swsc} 

\usepackage{amsmath}
\usepackage{graphicx}
\usepackage{txfonts}
\usepackage{subfigure}
\usepackage{epstopdf}
\usepackage{lineno}
\usepackage[authoryear,round]{natbib}
\usepackage[backref]{hyperref}

\newcommand{\eg}{{\it e.g.}}
\newcommand{\ie}{{\it i.e.}}

\hypersetup{colorlinks=true,citecolor=cyan,urlcolor=cyan,linkcolor=blue}

\begin{document}


\title{Flare Forecasting Using the Evolution of McIntosh Sunspot Classifications}

\titlerunning{Evolution-dependent McIntosh-Poisson Flare Forecasting}

\authorrunning{McCloskey et al.}

\author{
A. E. McCloskey\inst{1}, 
P. T. Gallagher\inst{1} 
\and 
D. S. Bloomfield\inst{2}
}

\institute{
School of Physics, Trinity College Dublin, College Green, Dublin 2, Ireland\\
\email{\href{mailto:mccloska@tcd.ie}{mccloska@tcd.ie}}
\and
Northumbria University, Newcastle upon Tyne, NE1~8ST, UK\\
\email{\href{mailto:shaun.bloomfield@northumbria.ac.uk }{shaun.bloomfield@northumbria.ac.uk }} 
}



\abstract
{
Most solar flares originate in sunspot groups, where magnetic field changes lead to energy build-up and release. 
However, few flare-forecasting methods use information of sunspot-group evolution, instead focusing on static point-in-time observations. 
Here, a new forecast method is presented based upon the 24-hr evolution in McIntosh classification of sunspot groups. Evolution-dependent $\geqslant$\,C1.0 and $\geqslant$\,M1.0 flaring rates are found from NOAA-numbered sunspot groups over December 1988 to June 1996 (Solar Cycle 22; SC22) before converting to probabilities assuming Poisson statistics. 
These flaring probabilities are used to generate operational forecasts for sunspot groups over July 1996 to December 2008 (SC23), with performance studied by verification metrics. 
Major findings are: i) considering Brier skill score (BSS) for $\geqslant$\,C1.0 flares, the evolution-dependent McIntosh-Poisson method ($\text{BSS}_{\text{evolution}}=0.09$) performs better than the static McIntosh-Poisson method ($\text{BSS}_{\text{static}} = -0.09$); 
ii) low BSS values arise partly from both methods over-forecasting SC23 flares from the SC22 rates, symptomatic of $\geqslant$\,C1.0 rates in SC23 being on average $\approx$\,80\% of those in SC22 (with $\geqslant$\,M1.0 being $\approx$\,50\%); 
iii) applying a bias-correction factor to reduce the SC22 rates used in forecasting SC23 flares yields modest improvement in skill relative to climatology for both methods ($\mathrm{BSS}^{\mathrm{corr}}_{\mathrm{static}} = 0.09$ and $\mathrm{BSS}^{\mathrm{corr}}_{\mathrm{evolution}} = 0.20$) and improved forecast reliability diagrams.
}

  
 

\keywords{
operational forecasting --
solar flares --
sunspot groups
}

\maketitle

\section{Introduction}

Solar flares are one of the most energetic space weather phenomena that affects the near-Earth environment. 
They most commonly originate within sunspot groups, where evolution of complex magnetic field leads to magnetic reconnection and subsequent large magnitudes of energy release. 
In the reconnection process, stored magnetic energy is rapidly converted to both thermal and kinetic energy in addition to non-thermal acceleration of particles \citep{Priest2002}. 
Solar flares, or coronal mass ejections (CMEs) if material is ejected, are understood to be caused by this magnetic reconnection process. 
Due to the high-energy radiation release and particle acceleration, these phenomena can have damaging effects on both Earth and space-based technologies (\eg, satellites and radio communication). 
Unlike CMEs that typically take 1--3 days to propagate to Earth after launch is detected, flare-related space weather impacts begin within minutes of flare onset \citep[\eg, ionospheric disturbances;][]{Mitra1974}. 
Therefore, it is of high priority that methods are developed to forecast when flares may occur, and the magnitude of energy release, in order to mitigate their effects.

Over the past several decades, there have been many published works focused on the classification of sunspot groups in terms of their complexity and their relation to flare production. 
The most well-known are the Mount Wilson \citep{Hale1919} and McIntosh \citep{McIntosh1990} schemes, classifying sunspots according to their magnetic and white-light structure, respectively. 
The relationship between these sunspot group classifications and flaring has been investigated in several studies and it was shown that the more ``complex'' sunspot-group classifications are associated with higher frequency and magnitude of flaring \citep{Waldmeier1947,Bornmann1994}.

In terms of solar flare prediction, one of the most established methods that has been developed to forecast solar flares is based upon sunspot-group classification, namely the McIntosh classification scheme. 
\citet{Gallagher2002} developed a Poisson-based method for calculating flare probabilities from the historical flaring rates of McIntosh classifications (publicly available at \href{http://www.solarmonitor.org}{www.solarmonitor.org}). 
Later this method was expanded upon and the performance of interpreting probabilities as dichotomous yes/no forecasts was verified by \citet{Bloomfield2012}, where it was shown that Poisson probabilities performed comparably to some of the more complex flare prediction methods in use at that time. 
There currently exists a vast quantity of prediction/forecasting methods including the most recent development of applying machine learning techniques to flare forecasting \citep[see, \eg,][]{Colak2009,Ahmed2013,Bobra2015}. 
For more information on the multitude of prediction/forecasting methods, see the recent comparison paper by \citet{Barnes2016} and references therein. 

Several space weather Regional Warning Centres (RWCs) make use of the Poisson-based flare forecasting approach. 
The US National Oceanographic and Atmospheric Administration (NOAA) Space Weather Prediction Centre (SWPC) RWC uses the McIntosh scheme as an input for their ``expert'' decision-rule system that is used to assign flaring probabilities to active regions \citep{McIntosh1990} that are augmented by experienced space weather forecasters prior to being issued. 
The UK Met Office Space Weather Operations Centre (MOSWOC) RWC also uses the historical flaring rates of McIntosh classes to calculate an initial forecast, again later adjusted by human forecasters. 
The performance of these operational forecasts have been evaluated and shown to perform well compared to more complex methods, with improvement in performance achieved by including the human editing of probabilities \citep{Crown2012,Murray2017}, also true for the Belgian Solar Influences Data Center (SIDC) RWC \citep{Devos2014}.

Until now, few forecasting methods account for evolution in sunspot-group properties, but there have been some research-focused studies considering evolution in sunspot-group classifications. 
\citet{Lee2012} investigated a subset of McIntosh classes alongside their 24-hr change in sunspot area, finding that groups which increased in area had a higher flaring rate compared to groups with  steady or decreasing area. 
Comparatively, \citet{McCloskey2016} calculated evolution-dependent flaring rates for the three components of the McIntosh classification scheme. 
It was shown that when sunspot groups evolve upward in their McIntosh class higher 24-hr flaring rates are observed, with lower flaring rates being true for downward evolution. 
So far, however, no verified forecasting methods have included the temporal evolution of sunspot-group classifications.

In this paper, we investigate the evolution of McIntosh sunspot-group classifications over 24-hr time scales as a method for forecasting solar flare magnitude and occurrence. 
The data we use is based upon \citet{McCloskey2016}, where historical flaring rates were calculated for McIntosh evolutions from the training period 1988--1996 (Solar Cycle 22; SC22), with more recent data from 1996--2008 (SC23) included for testing and forecast verification. 
In Section~\ref{sec:data_analysis} we provide more details on the data used in this study and the method used to produce flare probabilities. 
Section~\ref{sec:results} discusses the results of the forecasting method along with verification metrics and an exploration of the maximum performance possible when applying linear Cycle-to-Cycle rate corrections. 
Finally, in Section~\ref{sec:disc_conc} we present our conclusions and outlook for future work.

\section{Data Analysis} \label{sec:data_analysis}

\subsection{Data Sources} \label{subsec:data_sources}

The data used in this study are that analysed by \citet{McCloskey2016}, consisting of historical sunspot-group classifications and flare information collected by NOAA/SWPC. 
SWPC provide a daily Solar Region Summary (SRS) issued at 00:30 UT, with sunspot-group properties including NOAA active region number, heliographic coordinates, McIntosh and Mount Wilson classifications, and longitudinal extent. 
Additionally, solar flares associated with these regions were obtained from the Geostationary Operational Environment Satellite (GOES) event lists collated by SWPC. 
It is noted that the association of a flare to a specific NOAA active region is carried out by SWPC for up to three days after the event occurs. 
We chose to include all GOES 1\,--\,8\,\AA\ soft X-ray flares of C-class and above (\ie, $\geqslant$\,$10^{-6}$\,Wm$^{-2}$), with the reason for excluding flares below these magnitudes being the high background solar X-ray flux level at solar maximum that obscures B-class and lower flares.

The data used here as a training set for our forecasting method was taken from the SC22 period of 1 December 1988 to 31 July 1996, inclusive (Balch, 2011, private communication). 
It is noted that although SC22 is estimated to have commenced in September 1986 \citep{Hathaway1999}, the region-associated flare data from before December 1988 was not available and therefore could not be included here. 
This provided a data set of 24-hr flaring rates calculated for individual evolutions in McIntosh classification parameters, \ie, modified Zurich, penumbral or compactness classes.
However, it is important to note that in this study we chose to make use of the \emph{evolution in the full McIntosh classification} of each sunspot group rather than the evolution in the three separate components studied in \citet{McCloskey2016}. Section~\ref{subsec:fullmcint} outlines this distinction in further detail. 

The data used here as a test set was obtained from the publicly available NOAA/SWPC website (\href{ftp://ftp.swpc.noaa.gov/pub/warehouse/}{ftp://ftp.swpc.noaa.gov/pub/warehouse/}) over the SC23 period of 31 July 1996 to 13 December 2008, inclusive, in order to ensure an independent data set for forecast verification. 
Using the same method as \citet{McCloskey2016}, McIntosh classifications were extracted for each unique NOAA sunspot group along with the region-associated GOES X-ray flares. 
A total of 21,476 individual daily sunspot-group entries were extracted in the test period, corresponding to 3017 unique NOAA active regions. 
The total number of GOES soft X-ray flares associated with these regions was 8647, consisting of 7434 C-class, 1106 M-class, and 107 X-class flares. 

\subsection{Full McIntosh Classification Evolution} \label{subsec:fullmcint}

The McIntosh classification scheme is a long-established method for classifying the white-light structure of sunspot groups. 
It was first developed by \citet{Cortie1901} and later expanded upon and modified to include additional parameters \citep{Waldmeier1947,McIntosh1990}. 
The scheme is comprised of 17 different parameters which combine to form 60 different allowed classifications. 
These parameters are divided into three component classes: modified Zurich, penumbral and compactness (Zpc). 
In summary, `Z' describes the longitudinal extent of the sunspot group, `p' describes the size and symmetry of the penumbra of the leading spot and `c' describes the distribution of sunspots in the interior of the group. 
For a more detailed description of these components and their allowed combinations, see \citet{McIntosh1990}, \citet{Bornmann1994}, and \citet{McCloskey2016}.

Previously, it has been shown that the McIntosh classifications of sunspot groups and their flare productivity are related. 
Importantly, there is evidence that the McIntosh classification can capture differences in flaring rates for sunspot groups, with more complex classifications producing higher flaring rates overall \citep{Bornmann1994}. 
Building upon this, \citet{McCloskey2016} showed that the 24-hr evolution of McIntosh sunspot-group classifications show comparable results in terms of the rate of flare production -- sunspot groups that evolved upward in a classification component produced higher flaring rates, while downward evolution produced lower flaring rates. 
In this paper we make use of this statistical relationship to implement a method for flare forecasting using the 24-hr evolution of McIntosh classifications. 

As previously mentioned, instead of considering the evolution in only a single McIntosh component (\ie, $\mathrm{Z}_{1} \rightarrow \mathrm{Z}_{2}$ or $\mathrm{p}_{1} \rightarrow \mathrm{p}_{2}$ or $\mathrm{c}_{1} \rightarrow \mathrm{c}_{2}$), the full McIntosh class evolution of a sunspot group is extracted over 24 hours (\ie, $\{\mathrm{Zpc}\}_{1} \rightarrow \{\mathrm{Zpc}\}_{2}$). 
The main reasoning for this was to better capture the information in the evolution of the complete white-light structure of each sunspot group that was naturally excluded by considering only evolution in a single McIntosh component. 
Here, the average flaring rate associated with one unique $\{\mathrm{Zpc}\}_{1} \rightarrow \{\mathrm{Zpc}\}_{2}$ evolution is determined by extracting all instances of active regions that underwent that McIntosh class evolution. From this subset of active regions, the total number of flares that were produced within 24\,hr of that specific evolution are divided by the total number of regions in that subset.

\begin{figure}[t!]
\centering
\includegraphics[height=.75\textheight]{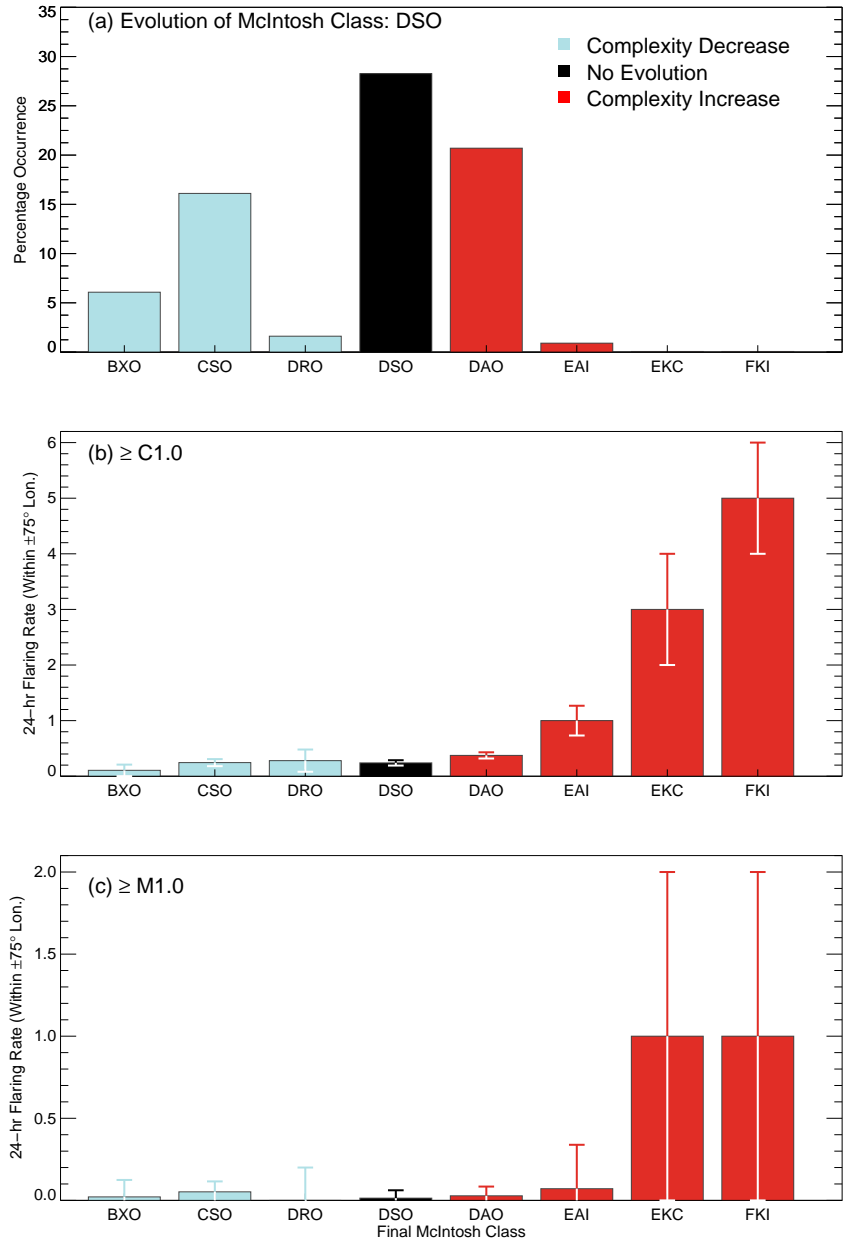}
\caption{
\label{fig:mcint_flrates}
Histograms showing the 24-hr evolution of sunspot groups starting as a DSO-type McIntosh classification (a), with bars representing the percentage of evolutions observed starting as DSO and evolving to a sub-group of McIntosh classifications. 
The corresponding evolution-dependent $\geqslant$\,C1.0 and $\geqslant$\,M1.0 flaring rates are shown in panels (b) and (c), respectively. 
Histogram bars are coloured by evolution: no evolution (black); upward evolution (dark red); downward evolution (light blue).
}
\end{figure}

To verify that the previously observed relationship between McIntosh-class evolution and flaring rate is also present when considering the full McIntosh classification, Figure~\ref{fig:mcint_flrates} depicts flaring rates for a selection of full McIntosh-class evolutions. 
This selection was chosen to represent evolution by evolving sequentially in at least one parameter (\eg, a DSO evolving to a BXO, followed by a DSO evolving to a CSO). 
Note that this graphical representation is less continuous to that shown in \citet{McCloskey2016}, since bars that lie two steps apart may depict evolution in two separate McIntosh components (rather than two steps in one component in the previous work). 
Figure~\ref{fig:mcint_flrates}a plots the occurrence-frequency distribution, with the most frequent occurrence once again being no evolution in McIntosh class over 24 hours (black bar). 
When evolution does occur, a DSO-type is most likely to evolve upward in penumbral class (\ie, to DAO) or downward in modified Zurich class (\ie, to CSO). 
This reflects the previous findings of \citet{McCloskey2016} where sunspot groups are most likely to remain the same classification and are not likely to evolve significantly over a 24-hr period (\ie, rarely more than two evolution steps in any one McIntosh component). 

Figure~\ref{fig:mcint_flrates}b displays the $\geqslant$\,C1.0 flaring rates associated with these selected McIntosh evolutions.  
This plot indicates that there are increasingly higher flaring rates associated with greater evolution steps upward in at least one McIntosh parameter, with the opposite true for greater evolution steps downward (\ie, sequentially decreasing rates). 
Additionally, for flaring rates $\lambda$, associated Poisson errors are calculated as $\Delta \lambda = 1/\sqrt{M}$, where $M$ is the total number of sunspot groups that underwent that evolution in McIntosh class. 
These are shown as error bars in both Figures~\ref{fig:mcint_flrates}b and \ref{fig:mcint_flrates}c, where the maximum error in flaring rate is $\pm$1. 
Similar behaviour is seen for $\geqslant$\,M1.0 flaring rates in Figure~\ref{fig:mcint_flrates}c, with higher flaring rates seen for evolution upward in McIntosh class, however due to low occurrence numbers these rates are deemed not statistically significant (\ie, $\lambda \pm \Delta \lambda$ encompasses zero). 
This relationship of McIntosh class evolution and flaring rates is comparable to the findings of \citet{McCloskey2016}. 

\subsection{Issuing Poisson Probabilities} \label{subsec:issuing_poisson_probs}

For the purpose of testing the forecast method in an operational manner, forecasts for $\geqslant$\,C1.0 and $\geqslant$\,M1.0 flares are issued for each 24-hr time window from 00:00\,UT in the form of probabilities of flare occurrence. 
It has been previously shown that the waiting-time distributions of soft X-ray flares from individual active regions is well represented by a time-dependent Poisson process with typical piece-wise constant flaring-rate timescales of $> 2-3$ days \citep{Wheatland2001}. As that work encompasses the full lifetime of individual active regions, and hence their evolution across McIntosh classes, we find the assumption of Poisson statistics suitable for our work. Here, we convert our evolution-dependent 24-hr flaring rates to probabilities as follows,
\begin{equation}
P_\lambda \left( N_{\mathrm{f}} \right) = \frac{\lambda^{N_{\mathrm{f}}}}{N_{\mathrm{f}}!} e^{-\lambda} \ ,
\end{equation}
where $N_{\mathrm{f}}$ is the number of flares expected to occur in a 24-hr period following an evolution and $\lambda$ is the average number of flares observed within the 24\,hr immediately following each unique evolution in McIntosh class. 
Note, these flaring probabilities are calculated separately for each unique full McIntosh evolution using the training set data of SC22. 
Hence, the probability of one or more flares occurring in a given time interval following an evolution is then calculated by,
\begin{equation}
P_\lambda \left( N_{\mathrm{f}} \geqslant 1 \right) = 1 - P_\lambda \left( N_{\mathrm{f}} = 0 \right) = 1 - e^{-\lambda} \ .
\end{equation}
By using a 24-hr flaring rate, the issued flaring probabilities are then valid for the following 24-hr period (\ie, 00:00\,UT to 00:00\,UT). 
Although the SWPC SRS files used to determine McIntosh-class evolution are issued at 00:30\,UT, here the forecast interval begins at 00:00\,UT as this is the end-time at which McIntosh classifications are constructed from the previous 24 hours. 

It is worth noting that there are certain circumstances where our evolution-dependent forecasting method will not be able to issue a forecast. 
This includes the first day a sunspot group appears on disk and therefore no evolution can have been observed, while there are a small number of full McIntosh-class evolutions that were not observed in the training data set and therefore no evolution-dependent flaring rate can be assigned in the test data set. 
Rather than disregard these sunspot groups from the analysis, we have chosen instead to use the standard static point-in-time flaring rates and hence probabilities for these cases based on the currently observed full McIntosh class. This satisfies the purpose of creating an operational forecasting method and allows for a more fair comparison of our evolution-dependent method with the original static method.

\section{Results} \label{sec:results}

\subsection{Forecast Verification} \label{subsec:verification}

Various verification metrics can be investigated to quantify the performance of a forecasting method. 
There are two main types of forecasting methods that are widely used, namely categorical and probabilistic. 
Dichotomous categorical forecasts have only two possible values when predicting if an event will occur (\ie, yes/no), whereas probabilistic forecasts yield a range of values (\ie, decimal percentage between 0 and 1). 
Here, we evaluate the performance of both the original static McIntosh method \citep{Gallagher2002} and our new evolution-dependent McIntosh method focusing on verification techniques suited for probabilistic forecasts. 
This allows for direct comparison of the two methods using probabilistic verification metrics that were not explored in the previous benchmarking study of \citet{Bloomfield2012}. 

One of the main quantities that assesses the performance of a  probabilistic forecast is the Brier score (BS). 
In its simplest form, BS is equivalent to the mean-squared error between the issued forecast probability, $f$ (\ie, 0--1), and the observed binary outcome for that forecast, $o$ (\ie, 0 or 1), 
\begin{equation} \label{eqn:BS}
\mathrm{BS} = \frac{1}{N} \sum_{i=1}^{N} (f_{i}-o_{i})^2 \ ,
\end{equation}
where $N$ is the total number of forecasts issued and $i$ identifies specific forecast-observation pairs. 
If the issued forecasts can be identified as groups of unique forecast probabilities, the BS can be decomposed into three components \citep{Murphy1973}, 
\begin{equation} \label{eq:bs_decomp}
\begin{split}
\mathrm{BS} &= \frac{1}{N} \sum_{k=1}^{K} n_{k} \left( f_{k} - \overline{o}_{k} \right)^2 - \frac{1}{N} \sum_{k=1}^{K} n_{k} \left( \overline{o}_{k} - \overline{o} \right)^2 + \overline{o} \left( 1 - \overline{o} \right) \ ,\\
            &= \mathrm{reliability} - \mathrm{resolution} + \mathrm{uncertainty} \ ,
\end{split}
\end{equation}
where $k$ identifies unique forecast-probability groups, $n_{k}$ is the number of occurrences in each $k$ group, $\overline{o}_{k}$ is the corresponding observed frequency of events in that $k$ group (\ie, the climatology for that unique forecast group) and $\overline{o}$ is the overall climatology of events for all valid forecast days. 
Climatology of events refers to the long-term average value of binary flare occurrence (\ie, 0 or 1) over the period of testing (\ie, SC23). 
Reliability is a measure of how close the issued probability of a unique forecast group is to the frequency of observed outcomes for that unique forecast group (\ie, the average binary outcome of their observed events), where a reliability value of 0 corresponds to a perfectly reliable forecast. 
The resolution term measures the difference between the climatology of the unique forecast groups and the overall climatology, which can be interpreted as the potential ability of the unique forecast groups to perform better than unskilled climatology (\ie, the higher the value of resolution the better). 
Finally, the uncertainty term measures the variability in the observed event frequency, which is independent of unique forecast grouping and is largest when an event is difficult to predict (\ie, occurring 50\% of the time) and smallest when an event occurs almost always or never.
In the context of this work, the issued forecast probabilities can be considered as binned into $k$ unique bins where each represents a unique McIntosh-class evolution (\eg, AXX to BXO). 

To interpret the performance of a forecast set, it is standard practice to normalise a verification metric score, $\mathrm{S}$, to that of a reference forecast, $\mathrm{S}_{\mathrm{ref}}$, by means of a skill score (SS),
\begin{equation} \label{eq:ss}
\mathrm{SS} = \frac{\mathrm{S} - \mathrm{S}_{\mathrm{ref}}}{\mathrm{S}_{\mathrm{perfect}} - \mathrm{S}_{\mathrm{ref}}} \ ,
\end{equation}
where $\mathrm{S}_{\mathrm{perfect}}$ is the score of a perfect forecast for the chosen verification metric. 
In the case of BS, a perfect forecast has a value of 0 and the reference forecast is typically taken to be that achieved by climatology, $\mathrm{BS}_{\mathrm{clim}}$ (equivalent to the uncertainty term in Equation~\ref{eq:bs_decomp}, as reliability and resolution cancel each other out). The Brier skill score (BSS) is then given as, 
\begin{equation} \label{eq:bss}
\mathrm{BSS} = \frac{\mathrm{BS} - \mathrm{BS}_{\mathrm{clim}}}{0 - \mathrm{BS}_{\mathrm{clim}}} = 1 - \frac{\mathrm{BS}}{\mathrm{BS}_{\mathrm{clim}}} \ .
\end{equation}
This can also be represented via the decomposed form of Equation~\ref{eq:bs_decomp} by the three components as,
\begin{equation} \label{eq:bss_decomp}
\mathrm{BSS} = 1 - \frac{\mathrm{reliability} - \mathrm{resolution} + \mathrm{uncertainty}}{\mathrm{uncertainty}} = \frac{\mathrm{resolution} - \mathrm{reliability}}{\mathrm{uncertainty}} \ .
\end{equation}
Table~\ref{t:bss} presents the three decomposed BS components and BSS for $\geqslant$\,C1.0 and $\geqslant$\,M1.0 flares for both the McIntosh static and evolution-dependent forecasting methods. 
Focusing on BSS values for $\geqslant$\,C1.0 flares, both methods achieve similar reliability values of 0.037 and 0.033, respectively. 
Considering now the resolution, as these values contribute to the overall BSS positively, if the value of resolution is greater than reliability the overall BSS will be positive. 
For the static method, despite being reasonably reliable it does not achieve a positive BSS ($-0.09$) as the value of resolution is too low (0.025) -- the climatology for many of the unique forecast groups are indistinguishable from the overall climatology (\ie, little forecast discrimination ability). 
Although the evolution-dependent method has a similar reliability value, its resolution (0.046) is higher, relative to both the static method and its own reliability term, contributing to a positive BSS (0.09). 
Achieving a positive value for BSS indicates that the evolution-dependent method is performing better than the climatology reference forecast, while the static method does not.

\begin{table}[t]
\caption{
Decomposed Brier score (BS) components and Brier skill score (BSS) for the McIntosh static and evolution-dependent forecast methods.
\label{t:bss}
}
\centering
\begin{tabular}{llcccr}
\hline\hline
Flaring           & Forecast  & \multicolumn{3}{c}{BS Components}      & BSS\\
Magnitude         & Method    & Reliability & Resolution & Uncertainty & \\
\hline
$\geqslant$\,C1.0 & Static    & 0.037       & 0.025      & 0.146       & $-$0.09\\
$\geqslant$\,C1.0 & Evolution & 0.033       & 0.046      & 0.146       & 0.09\\
$\geqslant$\,M1.0 & Static    & 0.017       & 0.003      & 0.038       & $-$0.36\\
$\geqslant$\,M1.0 & Evolution & 0.014       & 0.009      & 0.038       & $-$0.15\\
\hline
\end{tabular}
\end{table}

\begin{figure}
\centering
\includegraphics[width=\textwidth]{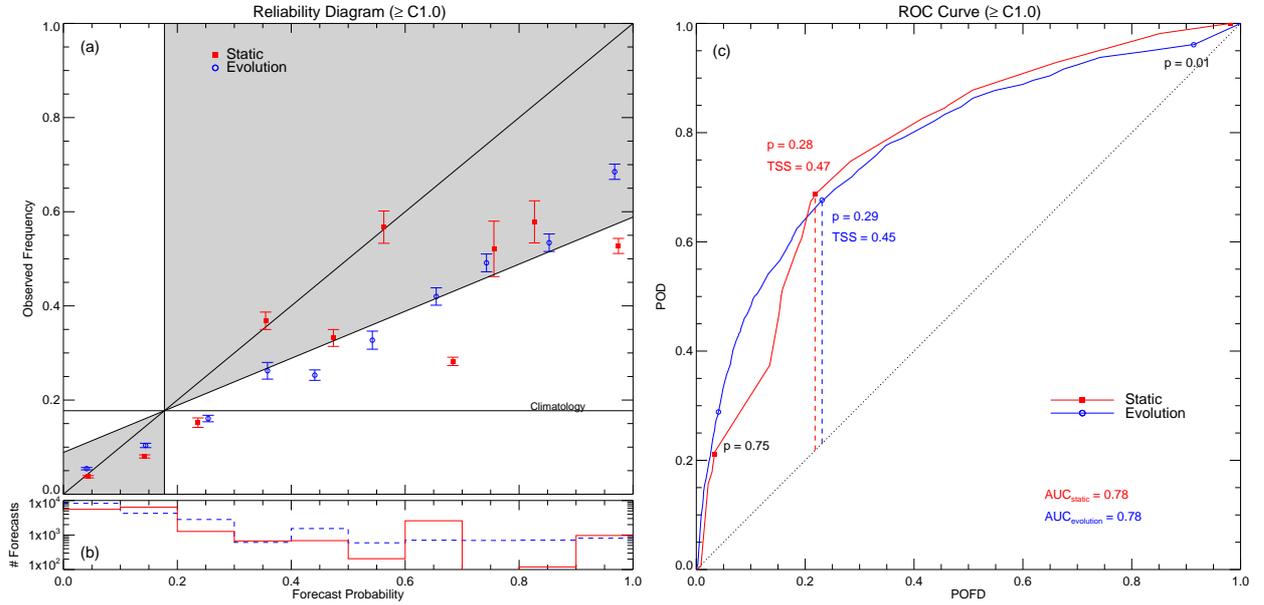}
\caption{
Reliability diagrams (panel a), sharpness (\ie, probability occurrence) plots (panel b), and ROC curves (panel c) for $\geqslant$\,C1.0 flares. Data for the McIntosh static forecast method are indicted by red filled squares (panels a and c) and solid histogram (panel b), while the evolution-dependent method is depicted by blue open circles (panels a and c) and dashed histogram (panel b).
\label{fig:rel_roc}
}
\end{figure}

In addition to skill scores, it is useful to visualise the performance of the forecast method. 
The two most popular visual diagnostics are reliability diagrams and relative operating characteristic (ROC) curves like those provided in Figures~\ref{fig:rel_roc}a and \ref{fig:rel_roc}c, respectively. 
Reliability diagrams indicate differences between forecast probabilities and the observed frequencies of events (similar to the reliability term of the BSS in Equation~\ref{eq:bs_decomp}), with forecast probabilities plotted along the horizontal axis, binned into sub-groups of forecasts, and the frequency of observed events for each sub-group plotted on the vertical axis. 
Here we chose to use 10\% probability intervals, $p$, with the associated Bayesian uncertainty for each bin shown as error bars, $\sigma_p = \sqrt{p(1-p)/(S+3)}$, where $S$ is the total number of forecast days in each probability bin \citep{Wheatland2005}, indicated in the sharpness plot of Figure~\ref{fig:rel_roc}b. 
The overall event climatology of events is plot as a horizontal and a vertical line, with the shaded area indicating the region that data contribute positively to BSS. 

Forecasts for the McIntosh static (red filled squares) and evolution-dependent (blue open circles) methods can be directly compared here, as both are applied to the same testing time period and so have the same climatology. 
For the static case, the majority of points lie within the shaded area, which can contribute positively to the BSS. 
However, while three points lie on the line of perfect reliability (\ie, $y=x$) most are found below this line, indicating the method is over-forecasting (\ie, the values of forecast probabilities are too high relative to the observed frequency of events for that forecast bin). 
It is interesting to note that the evolution-dependent case also appears to be over-forecasting, but in a more consistent manner (\ie, linearly biased from perfect reliability) than the static case. 
Notably, the static method achieves a worse (and negative) BSS compared to the evolution-dependent method, which is reflected in the reliability diagrams by more significant deviation of data points from the $y=x$ line and their relatively larger occurrence frequencies (\eg, for the static case, $p=0.6$--$0.7$ is the greatest outlier while being the third-most populated bin). 

For alternative verification purposes it is also possible to convert the probabilistic forecasts into dichotomous forecasts by probability thresholding. 
This is implemented by choosing a specific threshold and setting any forecast probability above that value to 1 (\ie, a `yes' forecast) and any forecast below it to 0 (\ie, a `no' forecast). 
The four possible arrangements of forecast-observation pairs can then be represented by a $2\times2$ contingency table consisting of: true positive forecasts (TP; hits), true negative forecasts (TN; correct rejections), false positive forecasts (FP; false alarms) and false negative forecasts (FN; missed flares). 
A ROC curve is then a visualisation of the probability of detection (also known as hit rate), $\mathrm{POD} = \mathrm{TP}/(\mathrm{TP} + \mathrm{FN}$), against the probability of false detection (also known as false alarm rate), $\mathrm{POFD} = \mathrm{FP}/(\mathrm{FP} + \mathrm{TN})$, as a function of probability threshold. 
A skillful forecast will have a higher success-ratio of events (POD) to failure-ratio of non-events (POFD), therefore the closer the curve is to the top left-hand corner the better. 
The ROC curve is also a visualisation of the True Skill Statistic ($\mathrm{TSS} = \mathrm{POD} - \mathrm{POFD}$), where the vertical distance of the curve above the diagonal line is the TSS value at that probability threshold (\ie, curves below the diagonal have negative TSS). 

Figure~\ref{fig:rel_roc}c displays the ROC curves for both the static (red filled squares) and evolution-dependent (blue open circles) methods, with probability thresholds of $p=0.01$ and 0.75 as well as the threshold probability corresponding to the maximum TSS value indicated for each method. 
Initially the ROC curves of both methods behave similarly, with marginally larger TSS for the static case. 
However, after the threshold probabilities that yield maximum TSS, noticeable divergence occurs with the evolution-dependent curve remaining relatively smooth until converging once again at higher probability thresholds. 
This is a direct result of the evolution-dependent method containing more forecasts with mid-to-high probabilities relative to the static method (\eg, the sharpness plot of Figure~\ref{fig:rel_roc}b). 
Furthermore, the area under the curve (AUC) is a measure of the accuracy of the forecast set, with areas of 1 corresponding to perfect forecasts and 0.5 corresponding to no-skill forecasts (indicated by the diagonal dashed line in Figure~\ref{fig:rel_roc}c). 
Both methods have AUC values of 0.78, indicating they have comparable dichotomous forecast accuracy when considering performance across the entire probability space. 
Equivalent figures for $\geqslant$\,M1.0 flares can be found in Appendix~\ref{app_m1}, showing qualitatively similar behaviour between the methods in terms of over-forecasting relative to the observed event frequency and similar values of AUC and maximum TSS. 

Considering the overall performance of the static and evolution-dependent methods, both appear to perform similarly when only considering their categorical forecast representation. 
However, with probabilistic verification metrics it becomes evident that the methods do not achieve the same level of performance. 
For BSS, the evolution-dependent method was shown to perform better in skill by a value of $\approx$\,0.2 when considering either $\geqslant$\,C1.0 or $\geqslant$\,M1.0 flares. 
In the decomposition of BS, while both methods achieve similar reliability values they differ in resolution, leading to better performance by the evolution-dependent method. 
In terms of optimising a forecasting method, it is possible to apply forecast-bias corrections to achieve more reliable forecasts. 
However, for those methods with unique forecast-probability groupings the resolution is fundamentally invariant to such corrections (\ie, with the sets of forecast-observation pairs remaining the same in each unique group, $\overline{o}_{k}$ and hence resolution in Equation~\ref{eq:bs_decomp} does not change). 
Considering that both methods are known to be over-forecasting (see Figure~\ref{fig:rel_roc}a), in Section~\ref{ssec:calib} we consider a basic bias correction to explore what the best performance of the methods could be in an ideal scenario. 

\subsection{Forecast-bias Correction} \label{ssec:calib}

\begin{figure}
\centering
\includegraphics[width=\textwidth]{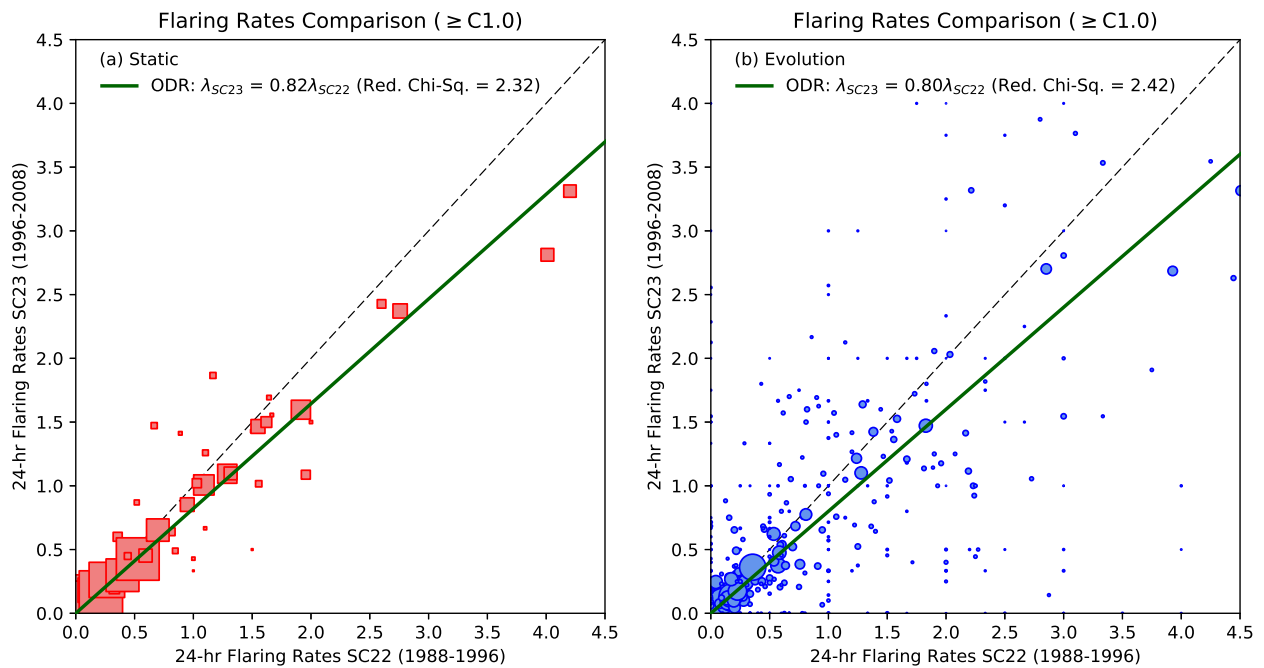}
\caption{
Comparison of $\geqslant$\,C1.0 24-hr flaring rates between SC22 (1988--1996) and SC23 (1996--2008) for the McIntosh static (panel a) and evolution-dependent (panel b) Poisson forecast methods. 
Dashed diagonal lines indicate the unity relation, while ODR best-fit linear relations are overlaid as thick lines. Best-fit slopes and reduced chi-squared values are also included.
\label{fig:flare_comp}
}
\end{figure}

Based on the results of verification performance for the static and evolution-dependent forecasting methods, we chose to investigate techniques to compensate for the over-forecasting of events in both cases. 
As both are Poisson-based methods derived from historical average flaring rates, the distributions of flaring rates were examined in the training (SC22) and test (SC23) data sets to investigate if a Cycle-to-Cycle variation existed.
Figure~\ref{fig:flare_comp} presents this comparison for $\geqslant$\,C1.0 flaring rates between SC22 (horizontal axes) and SC23 (vertical axes), for static (panel a) and evolution-dependent cases (panel b). 
The size of each data point corresponds to the total number of sunspot group occurrences, $M_{\mathrm{tot}} = M_{\mathrm{SC22}} + M_{\mathrm{SC23}}$, that are associated with each McIntosh class (panel a) or each evolution in full McIntosh class, such that larger data points were more frequently observed in both Solar Cycles. 

Considering the McIntosh static case in Figure~\ref{fig:flare_comp}a, 49 McIntosh classifications were observed in both the training and test data sets, while 518 full McIntosh-class evolutions were observed in both data sets (Figure~\ref{fig:flare_comp}b). 
These rate-rate plots were fit using the Orthogonal Distance Regression (ODR) method, as it takes account of uncertainties in both variables (\ie, $\Delta \lambda_\mathrm{SC22} = 1/\sqrt{M_\mathrm{SC22}}$ and $\Delta \lambda_\mathrm{SC23} = 1/\sqrt{M_\mathrm{SC23}}$). 
Fit intercepts were set to 0 to obtain slopes that can be later compared to rate-correction factors (RCFs) used to examine the possible influence of bias correction on forecast performance (see Section~\ref{ssec:optimisation}). 
Dashed diagonal lines in each panel indicate the unity slope (\ie, $\lambda_{\mathrm{SC23}} = \lambda_{\mathrm{SC22}}$), while ODR best-fit lines are displayed as thick lines. 
For the static method, the ODR best-fit is found (with a reduced chi-squared of 2.32) to be $\lambda_{\mathrm{SC23}} = (0.82\pm0.02)\lambda_{\mathrm{SC22}}$. 
As the fit slope is below unity, this indicates that the flaring rates for sunspot groups in the training period (SC22; 1988--1996) are on average higher than the those with the same McIntosh classifications in the test period (SC23; 1996--2008). 
For the evolution-dependent case, the same behaviour is found (\ie, $\lambda_{\mathrm{SC23}} = (0.80\pm0.02)\lambda_{\mathrm{SC22}}$ with a reduced chi-squared of 2.42). 
Given that the flaring rates deduced for both methods produce the same relationship within error, this indicates that the rate of flares produced by sunspot groups is Cycle-dependent. 
These differences in underlying flaring rates between training and testing periods directly contributes to over-forecasting by both methods when using the Poisson approach. 

Equivalent figures for $\geqslant$\,M1.0 flares can be found in Appendix~\ref{app_m1}. 
Qualitatively similar results are presented, but with even greater differences in flaring rates observed between SC22 and SC23 (\ie, $\lambda_{\mathrm{SC23}} = (0.52\pm0.02)\lambda_{\mathrm{SC22}}$ and $\lambda_{\mathrm{SC23}} = (0.49\pm0.02)\lambda_{\mathrm{SC22}}$ for the McIntosh static and evolution-dependent cases, respectively). 

\subsection{Forecast Performance Exploration}\label{ssec:optimisation}

As mentioned previously, it is possible to alter the performance of a forecasting method using bias-correction techniques. 
The results of the Cycle-to-Cycle flaring-rate comparison indicate that there is on average a difference in flaring rates for the same sunspot group type between the training and test data sets. 
Instead of relying solely on the best-fit slopes obtained from the rate-rate comparison, a range of RCFs were examined to find the optimum RCF conditioned on the BSS performance of the ``corrected'' forecasting methods. 
This technique works by adjusting the flaring rates obtained from the SC22 training period by multiplication with a RCF to produce new ``corrected'' flaring rates, with the standard Poisson approach once again applied to produce new ``corrected'' forecast probabilities. 

\begin{figure}[!t]
\centering
\includegraphics[height=.75\textheight]{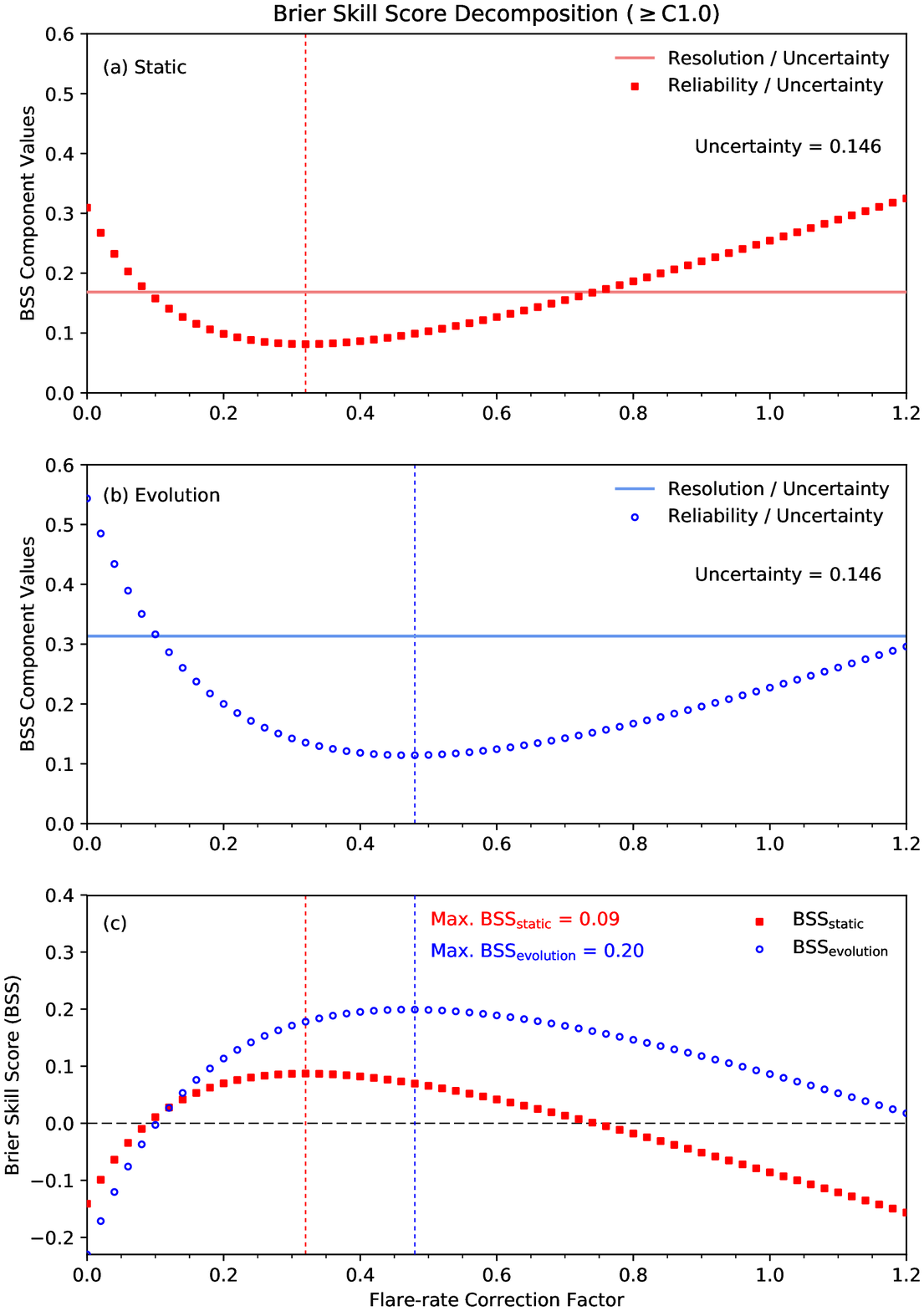}     
\caption{
Brier skill score (BSS) decomposition for the McIntosh static (panel a) and evolution-dependent (panel b) forecast methods for $\geqslant$\,C1.0 flares. 
BS components of reliability (data points), resolution (solid horizontal lines), and uncertainty (printed values) are displayed in panels a and b as a function of rate-correction factor (RCF) applied to the SC22 flaring rates. 
The resulting BSS is presented in panel c, also as a function of RCF applied to the SS22 flaring rates, for the static (red filled squares) and evolution-dependent (blue open circle) methods, with maximum values of BSS indicated by vertical dashed lines for both cases.
\label{fig:bss_decomp}
}
\end{figure}

The results of this analysis are presented in Figure~\ref{fig:bss_decomp}, showing the variation with RCF value of BSS and its components following the decomposition given in Equation~\ref{eq:bss_decomp}.  
Figure~\ref{fig:bss_decomp}a displays the variation of the resolution/uncertainty and reliability/uncertainty terms observed for the McIntosh static case, while the same for the evolution-dependent case is provided in Figure~\ref{fig:bss_decomp}b. 
The BSS-decomposed uncertainty term is constant (with a value of 0.146) and equal in both cases, as it only depends on the climatological frequency of events that is common to both methods. 
It is important to note that when using the decomposition of BSS correctly (\ie, when the forecast method comprises of distinctly unique forecast-probability groups), the resolution of the method is invariant under the bias correction performed by applying the RCF; evidenced by the normalised resolution term remaining constant as a function of RCF in both cases (\ie, horizontal lines). 
As the uncertainty-normalised reliability term is always positive and contributes negatively to BSS (see Equations~\ref{eq:bs_decomp} and \ref{eq:bss_decomp}), achieving the smallest possible value is highly desirable. 

For the McIntosh static method in Figure~\ref{fig:bss_decomp}a, the uncertainty-normalised reliability is optimized (\ie, minimized) at a value of 0.08 for a RCF of 0.32. 
Similarly for our evolution-dependent method, the minimum normalised reliability value of 0.11 is achieved for a RCF of 0.48 (Figure~\ref{fig:bss_decomp}b). 
For both cases this leads to the opposite behaviour for BSS as a function of RCF (Figure~\ref{fig:bss_decomp}c), with maximum BSS values of 0.09 and 0.20 achieved for the static and evolution-dependent methods, respectively. 
The optimal BS decomposed values and BSS are presented for $\geqslant$\,C1.0 and $\geqslant$\,M1.0 flares in Table~\ref{t:bss_corrected}. 
As mentioned before, the main difference between the two forecast methods is that our new evolution-based method achieves a resolution nearly twice that of the original static method, with uncertainty-normalised resolution values of 0.18 (static) and 0.31 (evolution-dependent). 
Optimising method reliabilities using a simple (admittedly \emph{post facto}) RCF technique as presented here leads to an approximately two-fold increase in BSS from the values in Table~\ref{t:bss}. 

\begin{table}[t]
\caption{
Optimized RCF-adjusted decomposed Brier score (BS) components and Brier skill score (BSS) for the McIntosh static and evolution-dependent forecast methods.
\label{t:bss_corrected}
}
\centering
\begin{tabular}{llccccr}
\hline\hline
Flaring           & Forecast  & Applied  & \multicolumn{3}{c}{BS Components}      & BSS\\
Magnitude         & Method    & SC22 RCF & Reliability & Resolution & Uncertainty & \\
\hline
$\geqslant$\,C1.0 & Static    & 0.32     & 0.012       & 0.025      & 0.146       & 0.09\\
$\geqslant$\,C1.0 & Evolution & 0.48     & 0.017       & 0.046      & 0.146       & 0.20\\
$\geqslant$\,M1.0 & Static    & 0.20     & 0.001       & 0.003      & 0.038       & 0.06\\
$\geqslant$\,M1.0 & Evolution & 0.30     & 0.005       & 0.009      & 0.038       & 0.09\\
\hline
\end{tabular}
\end{table}

\begin{figure}[t!] 
\centering
\includegraphics[ width=\textwidth]{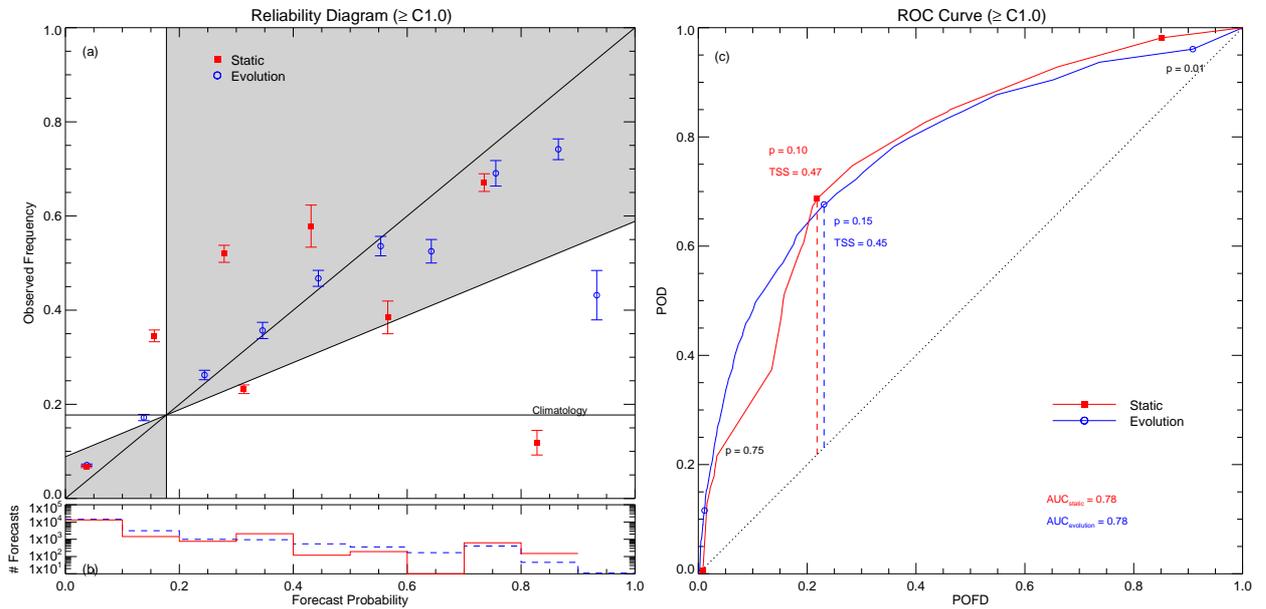}      
\caption{
As Figure~\ref{fig:rel_roc}, but using the BSS-optimised RCFs of 0.32 and 0.48 applied to the SC22 $\geqslant$\,C1.0 flaring rates for the McIntosh static and evolution-dependent forecast methods, respectively.
\label{fig:rel_roc_correct}
}
\end{figure}

``Corrected'' reliability diagrams and ROC curves are presented in Figure~\ref{fig:rel_roc_correct} using the optimized RCF values conditioned on maximising BSS to visualise the effect it has on forecast performance. 
The reliability diagrams of Figure~\ref{fig:rel_roc_correct}a confirm the McIntosh static (red filled squares) and evolution-dependent (blue open circles) forecast probabilities are both shifted to smaller values due to the RCFs applied being less than unity. 
Although this improves BSS for both methods, it does not appear to achieve a more reliable visual representation for the static method as several points appear to lie far from the line of perfect reliability (Figure~\ref{fig:rel_roc}a, red filled squares for comparison). 
In contrast, the evolution-dependent method appears to achieve a much more reliable visual representation than its equivalent uncorrected version (Figure~\ref{fig:rel_roc}a, blue open circles) with more points lying close to, or on, the line of perfect reliability. 
The ``corrected'' version of the ROC curves are presented in Figure~\ref{fig:rel_roc_correct}b, with no significant changes to the overall shape, area under the curve or maximum departure from the diagonal no-skill line. This is to be expected, as the application of the RCF only acts to shift the probability thresholds that the dichotomous categorical forecast statistics are calculated from (\ie, the forecast observation outcomes are unaltered). 
This could have implications for use in an operational situation: if bias-corrections are applied to create more reliable probabilistic forecasts, then the choice of probability threshold for evaluating the performance of subsequently-derived categorical metrics (or issuing of yes/no flare forecasts) needs to be reconsidered.

Equivalent plots for the RCF analysis and ``corrected'' reliability diagrams and ROC curves for $\geqslant$\,M1.0 flares can be found in Appendix~\ref{app_m1}, showing qualitatively similar results to the $\geqslant$\,C1.0 case (\ie, improvement in reliability and BSS).

\section{Discussion and Conclusion} \label{sec:disc_conc}

In this paper, we have examined the evolution of McIntosh sunspot group classifications and its application as a method for forecasting solar flares. 
Flaring rates calculated from sunspot-group evolution in McIntosh classifications during SC22 were used to produce probabilities for $\geqslant$\,C1.0 and $\geqslant$\,M1.0 flares within 24-hr forecast windows under the assumption of Poisson statistics. 
The reason for excluding flares below these magnitudes is the high background solar X-ray flux level at solar maximum that obscures B-class and lower flares. 
Additionally, due to the small number of X-class flares we chose to exclude the analysis of X-class and above as the large statistical errors lead to difficult interpretation of results. 
Similar to the results of \citet{McCloskey2016}, we find that upward evolution in at least one McIntosh classification component leads to higher flaring rates, with lower flaring rates occurring for downward evolution (Figure~\ref{fig:mcint_flrates}). 
Additionally, when sunspot groups evolve across multiple McIntosh classification components at the same time this behaviour is amplified -- \ie, increasingly higher (lower) flaring rates observed for greater upward (downward) evolution.

These flaring rates were converted to Poisson probabilities and applied to an independent test data set from SC23 to assess forecast performance, both for the original static point-in-time McIntosh forecasting method and our new evolution-dependent method. 
BSS was calculated for both, with the evolution-dependent method achieving a positive value for $\geqslant$\,C1.0 flares ($\mathrm{BSS}_{\mathrm{evolution}} = 0.09$), indicating that its performance surpasses that of climatology. 
In contrast, the static method performed worse than climatology ($\mathrm{BSS}_{\mathrm{static}} = -0.09$). 
Importantly, the determining factor for the difference in performance is that the evolution-dependent method achieves greater resolution than its static counterpart, indicating that the observed event occurrence averaged across the individual full-McIntosh class evolutions (\ie, unique forecast probability groups in the decomposed form of BS) is more separated from climatology than the same quantity averaged across individual static McIntosh classes. 
For $\geqslant$\,M1.0 flares the evolution-dependent method again performs better than the static method, but as both BSS values are negative ($\mathrm{BSS}_{\mathrm{evolution}} = -0.15$ and $\mathrm{BSS}_{\mathrm{static}} = -0.36$) this indicates that they do not perform as well as climatology. 
Reliability diagrams and ROC curves were also investigated, with a bias of over-forecasting clear in both methods (Figures~\ref{fig:rel_roc}a and \ref{fig:rel_roc}b). 

This tendency to over-forecast was investigated by comparing the flaring rates for the training data from SC22 with those of the test data from SC23 using an ODR technique to fit the rate-rate relations. 
Considering previous studies, it has been shown that the level of activity in SC23 is lower compared to earlier Cycles. 
For example, \citet{Joshi2005} report that the number of H$\alpha$ flare events was lower in SC23 compared to SC21 and SC22, while \citet{Joshi2015} found that there was a significant decrease in the total soft X-ray flare index (a measure of flare activity) in SC23 compared to SC21 and SC22.
These results agree well with our finding SC23 rates being $\approx$80\% and $\approx$50\% of those in SC22 for $\geqslant$\,C1.0 and $\geqslant$\,M1.0 flares, respectively (Figures~\ref{fig:flare_comp} and \ref{fig:flare_comp_m1}).

To explore the maximum-achievable performance by the McIntosh-Poisson forecasting methods, a range of rate-correction factors (RCFs) were explored through application to the original SC22 flaring rates to bias-correct the forecast probabilities. 
The optimal value of RCF for $\geqslant$\,C1.0 flares (\ie, that achieving maximum BSS) was found to be 0.32 for the static method, while the evolution-dependent method has a weaker correction factor of 0.48 (Figure~\ref{fig:bss_decomp}). 
Interestingly, these RCFs differ from the Cycle-to-Cycle ODR linear rate-rate slopes of $\approx$\,0.80, although the ODR-determined value is admittedly obtained with no information feeding back from the application of the adjusted flaring rates in forecasting. 
The resulting maximum values for corrected BSS were found to be 0.09 and 0.20 for the static and evolution-dependent methods, respectively. 
These correspond to a two-fold increase in BSS that confirms the lowering of forecast probabilities issued for SC23 yields better performance for both methods, evidenced by improved reliability diagrams (Figure~\ref{fig:rel_roc_correct}a). 
To put these values in context, \citet{Barnes2016} compared several flare-forecasting methods using standard verification metrics to assess performance. 
To ensure direct comparison of the methods, a common data set was used where all methods issued forecasts for each data entry, analogous to daily operational flare forecasts and therefore the most suitable for comparing to the operational methods presented here. 
The maximum BSS achieved for $\geqslant$\,C1.0 flares in 24-hr forecast windows by any of the methods in \citet{Barnes2016} was 0.32 (see their Table 4). 
After optimal bias-correction was determined and applied, our McIntosh evolution-dependent method achieved a BSS approaching but still less than this (\ie, $\mathrm{BSS}_{\mathrm{evolution}}^{\mathrm{corr}} = 0.20$). 

It is noted that the bias-correction method applied here determines the systematic differences in flaring rates between training and test periods from \emph{post facto} analysis. 
To be truly operational, the application of pre-forecast bias correction requires prior knowledge of these differences in rates. 
Therefore, predictions for the next Solar Cycle could provide the bias-correcting RCF for the next forecast test period.


\begin{acknowledgements}
The authors thank Dr Chris Balch (NOAA/SWPC) for providing the 1988--1996 data used in this research and Dr Graham Barnes (NWRA/CoRA) for useful discussions on Brier score decomposition. 
A.E.McC. is supported by an Irish Research Council Government of Ireland Postgraduate Scholarship and D.S.B. is supported by the European Union Horizon 2020 research and innovation programme under grant agreement No.~640216 (FLARECAST project; \href{http://flarecast.eu}{flarecast.eu}). The editor thanks two anonymous referees for their assistance in evaluating this paper.
\end{acknowledgements}

\appendix

\section{Forecast Verification of Flares At/Above M1.0} \label{app_m1}

\begin{figure}[t!]
\centering
\includegraphics[width=\textwidth]{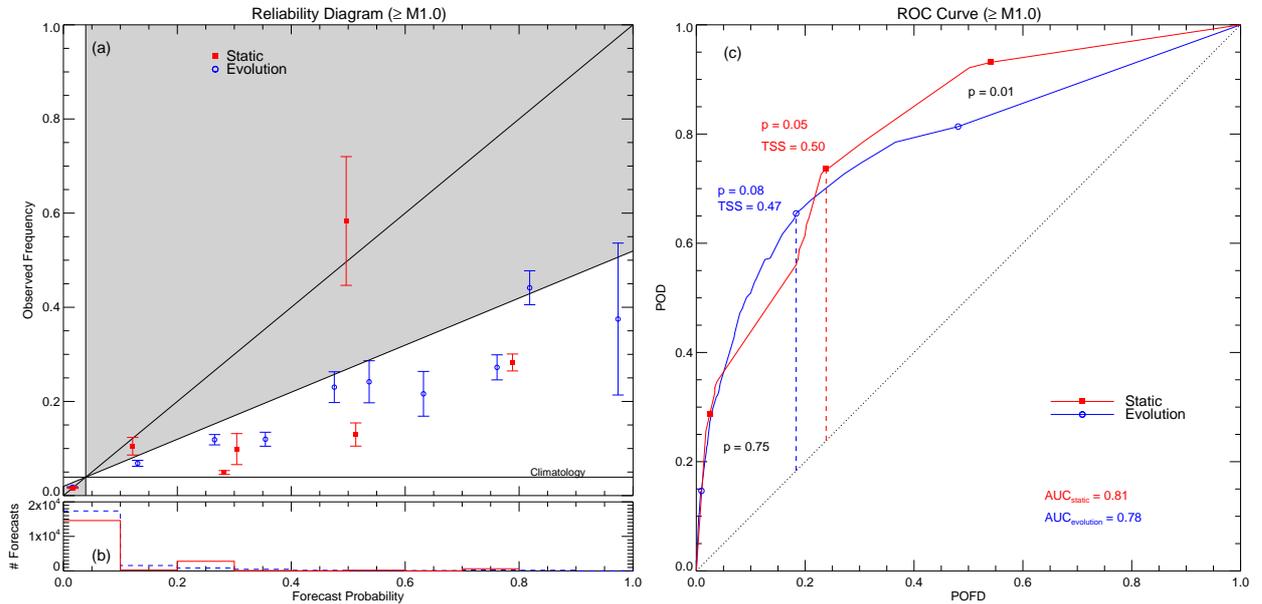}
\caption{
Reliability diagrams (panel a), sharpness (\ie, probability occurrence) plots (panel b), and ROC curves (panel c) for $\geqslant$\,M1.0 flares. Data for the McIntosh static forecast method are indicted by red filled squares (panels a and c) and solid histogram (panel b), while the evolution-dependent method is depicted by blue open circles (panels a and c) and dashed histogram (panel b).
\label{fig:rel_roc_m1}
}
\end{figure}

\begin{figure}[ht!]
\centering
\includegraphics[width=\textwidth]{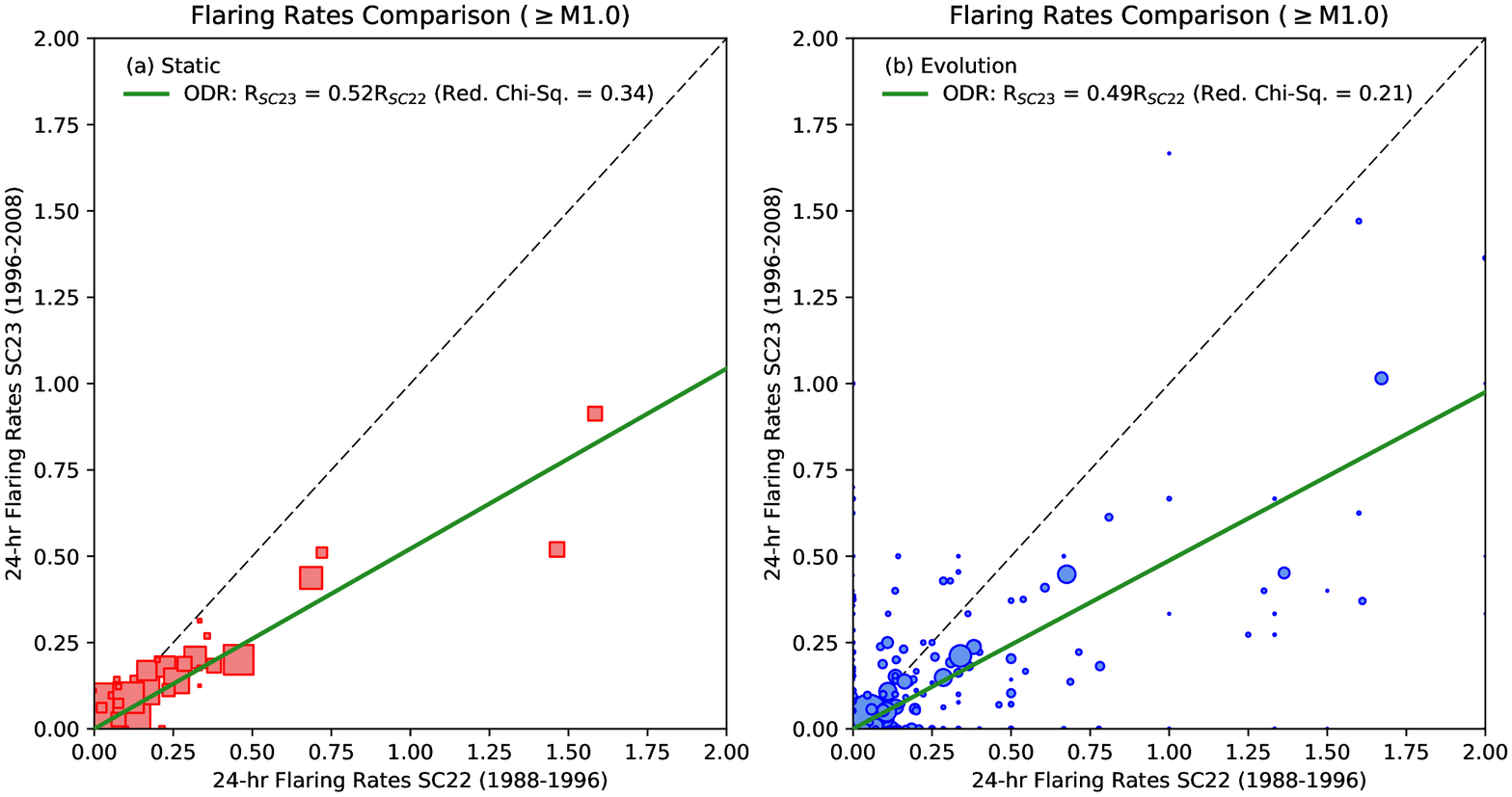}
\caption{
Comparison of $\geqslant$\,M1.0 24-hr flaring rates between SC22 (1988--1996) and SC23 (1996--2008) for the McIntosh static (panel a) and evolution-dependent (panel b) Poisson forecast methods. 
Dashed diagonal lines indicate the unity relation, while ODR best-fit linear relations are overlaid as thick lines. Best-fit slopes and reduced chi-squared values are also included.
\label{fig:flare_comp_m1}
}
\end{figure}

\begin{figure}[ht!]
\centering
\includegraphics[height=.75\textheight]{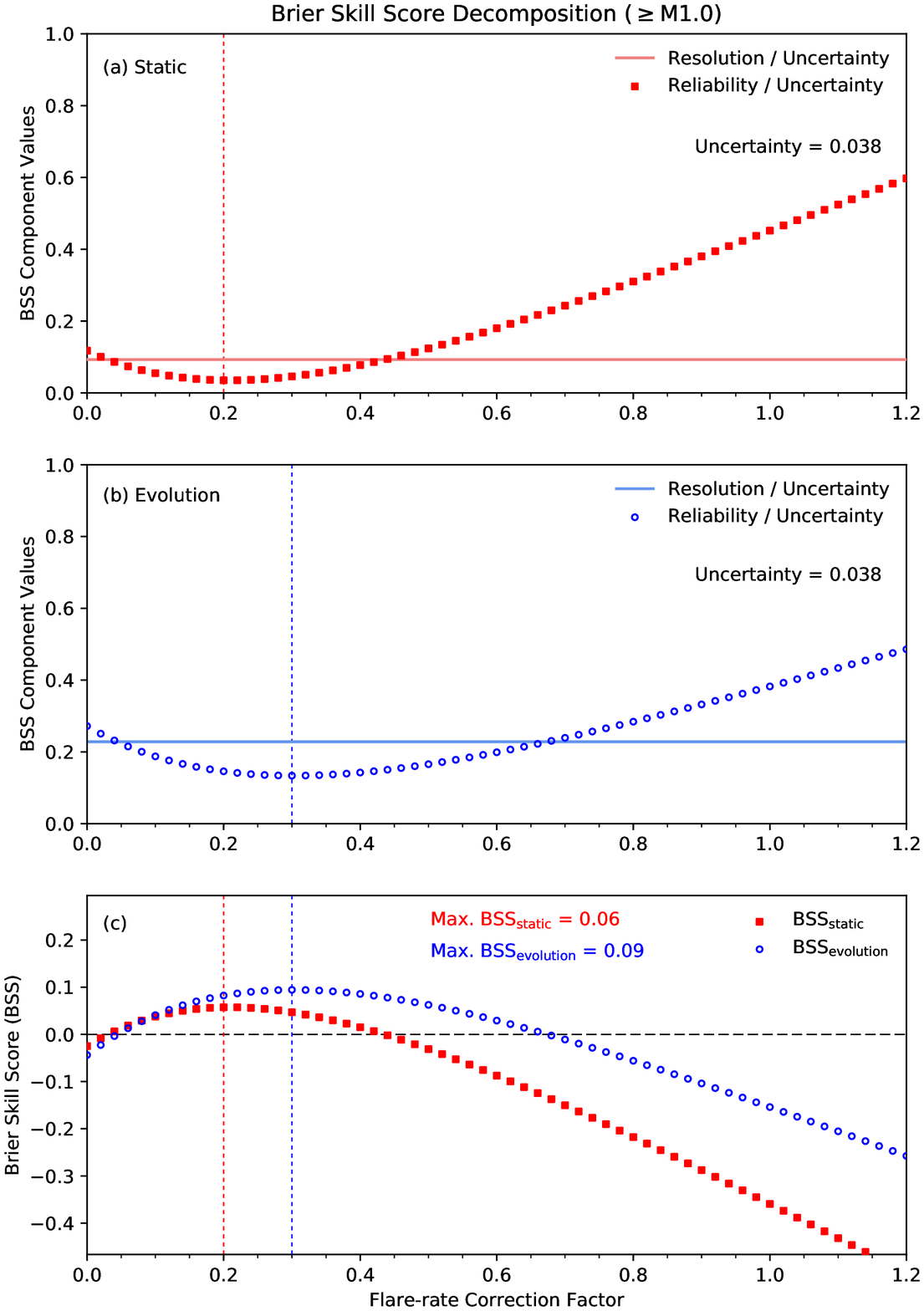}     
\caption{
Brier skill score (BSS) decomposition for the McIntosh static (panel a) and evolution-dependent (panel b) forecast methods for $\geqslant$\,M1.0 flares. 
BS components of reliability (data points), resolution (solid horizontal lines), and uncertainty (printed values) are displayed in panels a and b as a function of rate-correction factor (RCF) applied to the SC22 flaring rates. 
The resulting BSS is presented in panel c, also as a function of RCF applied to the SS22 flaring rates, for the static (red filled squares) and evolution-dependent (blue open circle) methods, with maximum values of BSS indicated by vertical dashed lines for both cases.
\label{fig:bss_decomp_m1}
}
\end{figure}

\begin{figure}[ht!]
\centering
\includegraphics[ width=\textwidth]{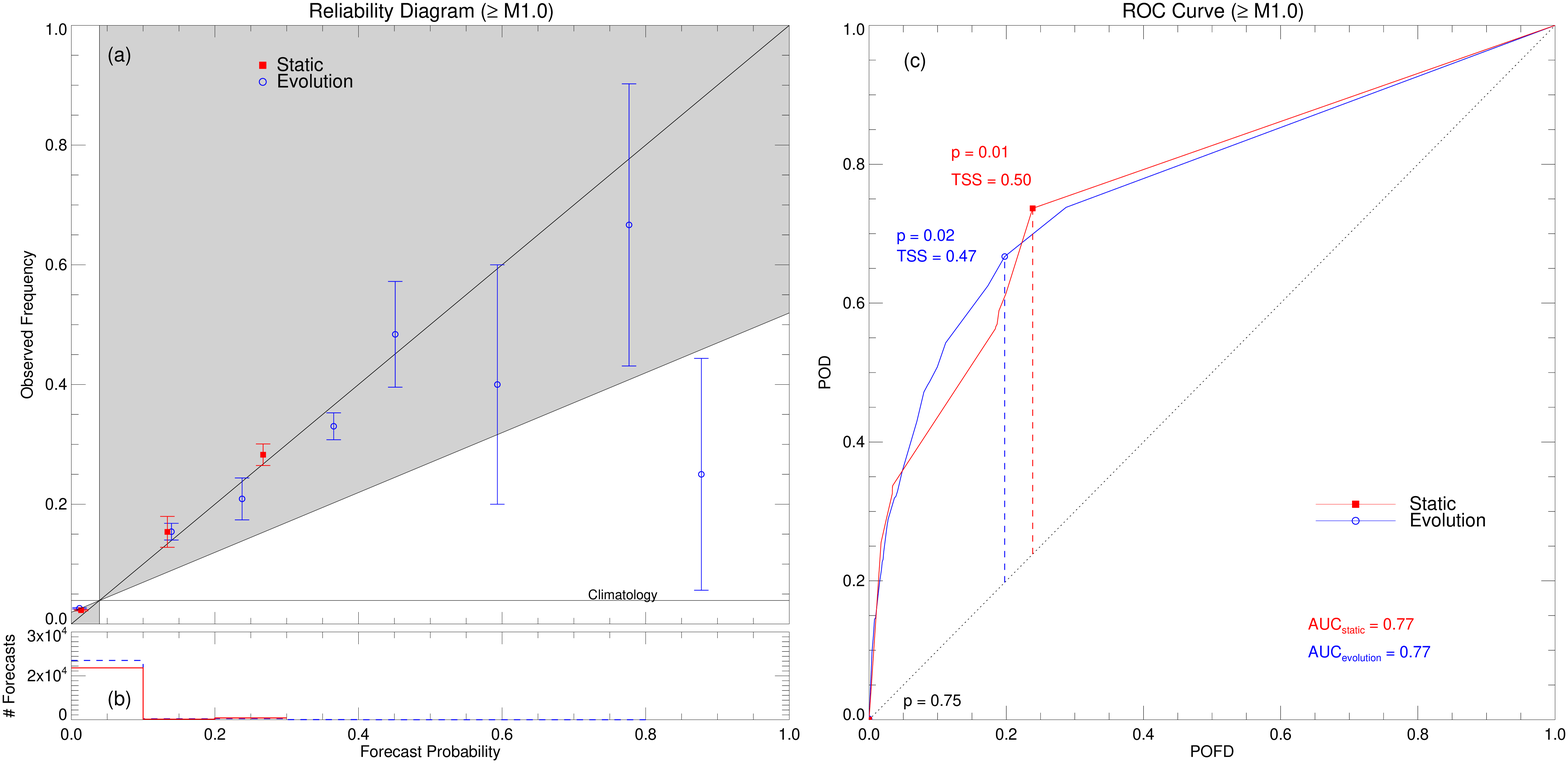}     
\caption{
As Figure~\ref{fig:rel_roc_m1}, but using the BSS-optimised RCFs of 0.20 and 0.30 applied to the SC22 $\geqslant$\,M1.0 flaring rates for the McIntosh static and evolution-dependent forecast methods, respectively.
\label{fig:rel_roc_correct_m1}
}
\end{figure}

Here we present equivalent figures to those in Section~\ref{sec:results}, but for $\geqslant$\,M1.0 flares. 
Reliability diagrams and ROC curves (equivalent to Figure~\ref{fig:rel_roc}) are plotted in Figure~\ref{fig:rel_roc_m1}. 
Following from this, the flare rate comparison between SC22 and SC23 (equivalent to Figure~\ref{fig:flare_comp}) is shown in Figure~\ref{fig:flare_comp_m1}. 
Finally, the BSS decomposition as a function of RCF and the ``corrected'' reliability diagrams and ROC curves are provided in Figures~\ref{fig:bss_decomp_m1} and \ref{fig:rel_roc_correct_m1}, respectively (equivalent to Figures~\ref{fig:bss_decomp} and \ref{fig:rel_roc_correct}).

\newpage
\clearpage

\bibliography{FF_Paper}


\end{document}